\documentclass{iopart}

\usepackage{iopams}
\usepackage{epsf}
\usepackage{cite,fleqn}

\newcommand{\bd}{\begin{displaymath}}
\newcommand{\ed}{\end{displaymath}}
\newcommand{\be}{\begin{equation}}
\newcommand{\ee}{\end{equation}}
\newcommand{\ba}{\begin{eqnarray}}
\newcommand{\ea}{\end{eqnarray}}

\begin{document}

\title[Evolution of the wave function in a three-grating interferometer]
{Evolution of the wave function of an atom hit by a photon in a
three-grating interferometer}

\author{D Arsenovi\'c$^1$, M Bo\v zi\'c$^1$, A S Sanz$^2$ and
M Davidovi\'c$^3$}

\address{$^1$Institute of Physics, University of Belgrade,\\
Pregrevica 118, 11080 Belgrade, Serbia\\
$^2$Instituto de F\'{\i}sica Fundamental,\\
Consejo Superior de Investigaciones Cient\'{\i}ficas,\\
Serrano 123, 28006 Madrid, Spain\\
$^3$Faculty of Civil Engineering, University of Belgrade,\\
Bulevar Kralja Aleksandra 73, 11000 Belgrade, Serbia}

\eads{\mailto{arsenovic@phy.bg.ac.yu},\mailto{bozic@phy.bg.ac.yu},
\mailto{asanz@imaff.cfmac.csic.es},\mailto{milena@grf.bg.ac.yu}}

\begin{abstract}
In 1995, Chapman \etal (1995 {\it Phys.\ Rev.\ Lett.}\ {\bf 75}
2783) showed experimentally that the interference contrast in
a three-grating atom interferometer does not vanish under the presence
of scattering events with photons, as required by the complementarity
principle.
In this work we provide an analytical study of this experiment,
determining the evolution of the atom wave function along the
three-grating Mach-Zehnder interferometer under the assumption that
the atom is hit by a photon after passing through the first grating.
The consideration of a transverse wave function in momentum
representation is essential in this study.
As is shown, the number of atoms transmitted through the third grating
is given by a simple periodic function of the lateral shift along this
grating, both in the absence and in the presence of photon scattering.
Moreover, the relative contrast (laser on/laser off) is shown to be a
simple analytical function of the ratio $d_p/\lambda_i$, where $d_p$
is the distance between atomic paths at the scattering locus and
$\lambda_i$ the scattered photon wavelength.
We argue that this dependence, being in agreement with experimental
results, can be regarded to show compatibility of the wave and
corpuscle properties of atoms.
\end{abstract}






\section{Introduction}
\label{sec1}

\begin{figure}
 \begin{center}
 \epsfxsize=12cm {\epsfbox{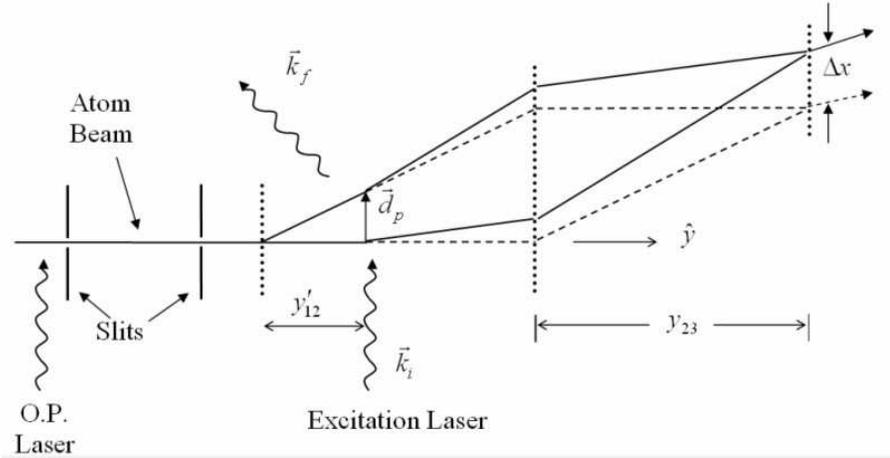}}
 \caption{\label{fig1}
  Sketch of the experimental three-grating interferometer used by
  Chapman \etal~\cite{chapman1,chapman2}.}
 \end{center}
\end{figure}

In an experiment performed by Chapman \etal \cite{chapman1} in
1995, single photons were scattered off atoms which passed through the
first grating of a three-grating Mach-Zehnder interferometer
\cite{chapman2}.
The purpose of this experiment was to study the influence of
photon scattering events on the atom interference.
The dependence of the atom transmission through the third grating on
the distance $y'_{12}$ between the place where the scattering event
occurred and the first grating (figure~\ref{fig1}) was then
investigated.
For each value of $y'_{12}$, the transmission was measured as a
function of the lateral shift $\Delta x_3$ of the third grating,
showing that the relative fringe contrast of the transmission depended
on the ratio $d_p/\lambda_i$, where $\lambda_i$ is the scattered
photon wavelength, and $d_p = y'_{12} \lambda /d$ is the distance
between two atomic paths at the scattering locus; in the latter
relation $d$ is the grating constant, $\lambda = h/mv = 2\pi/k$ is
the atomic de Broglie wavelength, and $v$ and $k$ are atomic initial
velocity and wave number, respectively.

The experiment showed that the contrast decreases to zero for
$d/\lambda_i \approx 0.5$, and several revivals with decreasing
relative maxima follow as increases \cite{chapman1,chapman2}.
Chapman \etal associated the loss of coherence with
complementarity and the subsequent revival with the spatial
resolution function of a single scattered photon.
Moreover, they also considered that their experiment addresses the
questions: Where the coherence is lost and how it might be regained?
These questions, in particular revivals of contrast, have been the
subject of discussions and studies \cite{drezet,pfau,guo,eberly}.

Here, we propose an explanation for the experimental results observed
by Chapman \etal \cite{chapman1} by determining the evolution
of the wave function of an atom in a three-grating interferometer in
two cases: a) the atom moves freely between the gratings and b) the
atom is hit by a photon between the first and second grating.
The consideration of a transverse wave function in momentum
representation is essential in our explanation.

\section{Evolution of the wave diffracted by a grating}
\label{sec2}

Consider an initial stationary atomic monochromatic wave, spreading
along the $y$-axis, that strikes a one-dimensional grating parallel
to the $x$-axis at $y = 0$,
\be
 \Psi(x,y,t) = e^{-i\omega t} \psi^i (x,y)
  = B^i e^{-i\omega t} e^{iky} , \qquad y < 0 ,
 \label{eq1}
\ee
where $B^i$ is a constant.
After reaching the grating, this incident wave is being transformed
into
\ba
 \Psi (x,y,t) & = & e^{-i\omega t} \psi(x,y) , \qquad y \ge 0 ,
 \label{eq2} \\
 \psi(x,y) & = & \frac{e^{iky}}{\sqrt{2\pi}} \ \!
  \int_{-\infty}^\infty dk_x \ \! c(k_x) \ \!
  e^{ik_x x} e^{-ik_x^2 y/2k} , \qquad y \ge 0 .
 \label{eq3}
\ea
Here, we consider gratings such that the function $c(k_x)$ has non
negligible values only for $k^2 \gg k_x^2$ \cite{bozic1,bozic2}.
Under this assumption, satisfies the Helmholtz equation.
The function $c(k_x)$ gives the probability amplitude of transverse
momenta and is determined by the boundary conditions at the grating.
If the grating is completely transparent inside the slits (union of
slit areas is denoted by $A$) and completely absorbing outside them,
$c(k_x)$ is given by the following equation \cite{bozic1,bozic2}:
\be
 c(k_x) = \frac{1}{\sqrt{2\pi}} \ \!
  \int_{-\infty}^\infty dx' \psi (x',0^+) e^{-ik_x x'}
  = \frac{1}{\sqrt{2\pi}} \ \!
  \int_A dx' \psi^i (x',0^-) e^{-ik_x x'} ,
 \label{eq4}
\ee
where $\psi (x',0^+)$ is the wave function just behind the first
grating and $\psi^i (x',0^-)$ is the wave function just before the
first grating.

As shown by Arsenovi\'c \etal~\cite{bozic3}, the solution
of the Helmholtz equation, $\psi(x,y)$, given by (\ref{eq3}),
is equivalent to the Fresnel-Kirchhoff solution
\be
 \psi(x,y) = \sqrt{\frac{k}{2\pi y}} \ \! e^{-i\pi/4} e^{iky}
  \int_{-\infty}^\infty dx' \psi(x',0^+) e^{ik(x-x')^2/2y} .
 \label{eq5}
\ee
The latter form is very useful because one can easily show from it that
there exists direct proportionality between the functions $\psi(x,y)$
and $c(kx/y)$ in the region far from the grating:
\be
 \psi(x,y) = \sqrt{\frac{k}{y}} \ \! e^{-i\pi/4} e^{ikx^2/2y}
  c(kx/y) e^{iky} .
 \label{eq6}
\ee
The solution given in (\ref{eq2}) and (\ref{eq3}) suggests that,
behind the grating, the atom continues propagating with the initial
longitudinal momentum, since a change of it is negligible.
However, there is a probability density $|c(k_x)|^2$ that an atom
acquires a transverse momentum $p_x = \hbar k_x$.
This justifies \cite{pfau,guo} the substitution of $y$ by $\hbar kt/m$
in the integrand of (\ref{eq3}) and defining the so-called {\it wave
function of the transverse motion},
\ba
 \psi^{tr}(x,t=ym/\hbar k) & = & \frac{1}{\sqrt{2\pi}} \ \!
  \int_{-\infty}^\infty dk_x c(k_x,t) e^{ik_x x}
  \nonumber \\
  & = & \frac{1}{\sqrt{2\pi}} \ \!
  \int_{-\infty}^\infty dk_x c(k_x) e^{-ik_x^2\hbar t/2m} e^{ik_x x} ,
 \label{eq7}
\ea
where $c(k_x,t)$ is the time-dependent transverse wave function in
momentum representation,
\be
 c(k_x,t) = \frac{1}{\sqrt{2\pi}} \ \!
  \int_{-\infty}^\infty dx \psi^{tr} (x,t) e^{-ik_x x}
  = c(k_x) e^{-ik_x^2\hbar t/2m} .
 \label{eq8}
\ee
As can be seen, $\psi^{tr}(x,t)$ has the form of a non-stationary
solution of the one-dimensional free-particle time-dependent
Schr\"odinger equation.
The solution (\ref{eq2}) is then a product \cite{bozic1,bozic2,bozic3}
of a longitudinal plane wave and a non-stationary transverse wave
function,
\be
 \Psi (x,y,t) = e^{-i\omega t} e^{iky} \psi^{tr}(x,t) .
 \label{eq9}
\ee


\section{Evolution of the diffracted wave after the atom is hit by a
photon}
\label{sec3}

We shall now use the above atomic wave function behind the grating and
its interpretation to determine the atomic wave function after the atom
absorbed and reemitted a photon somewhere along the $x$-axis at a time
$t'_{12}$ and a distance $y'_{12} = vt'_{12} = (\hbar k/m) t'_{12}$
from the first grating.
As a result of the scattering with the photon, there is a change of the
atomic transverse momentum $\Delta k_x$, which also leads to the change
of the wave function in the momentum representation.
We denote the wave function after the photon-atom scattering event in
momentum representation as $c_{\Delta k_x} (k_x,t)$.
It has to satisfy
\be
 \left\arrowvert c_{\Delta k_x} (k_x,t'_{12}) \right\arrowvert^2 =
  \left\arrowvert c(k_x - \Delta k_x,t'_{12}) \right\arrowvert^2 .
 \label{eq10}
\ee
From this relation, it follows
\be
 c_{\Delta k_x} (k_x,t'_{12}) =
  c(k_x - \Delta k_x,t'_{12}) e^{if(\Delta k_x, k_x)} ,
 \label{eq11}
\ee
where $f(\Delta k_x, k_x)$ is (for now) an unknown phase function.
The corresponding transverse wave function at time $t'_{12}$ is then
given by:
\be
 \psi^{tr}_{\Delta k_x}(x,t'_{12}) = \frac{1}{\sqrt{2\pi}} \ \!
  \int_{-\infty}^\infty dk_x c_{\Delta k_x}(k_x,t'_{12}) e^{ik_x x} ,
 \label{eq12}
\ee
which should satisfy
\be
 \left\arrowvert \psi^{tr}_{\Delta k_x}(x,t'_{12}) \right\arrowvert^2
  = \left\arrowvert \psi^{tr}(x,t'_{12}) \right\arrowvert^2 .
 \label{eq13}
\ee
Using (\ref{eq11}), one shows that the latter condition is fulfilled if
\be
 f(\Delta k_x, k_x) = 0 .
 \label{eq14}
\ee
After substitution of (\ref{eq11}) and (\ref{eq14}) into (\ref{eq12}),
one finds that, just after photon-atom scattering event, the atomic
wave function becomes:
\ba
 \psi^{tr}_{\Delta k_x}(x,t'_{12}) & = & \frac{1}{\sqrt{2\pi}} \ \!
  e^{-i\Delta k_x^2\hbar t'_{12}/2m} \nonumber \\
  & & \times
  \int_{-\infty}^\infty dk_x c(k_x - \Delta k_x)
   e^{-i k_x^2\hbar t'_{12}/2m} e^{ik_x (x + \Delta x_0)} ,
 \label{eq15}
\ea
where we have introduced the magnitude
\be
 \Delta x_0 = \frac{\Delta k_x \hbar t'_{12}}{m}
  = \frac{\Delta k_x y'_{12}}{k} .
 \label{eq16}
\ee
Assuming the function (\ref{eq15}) keeps the same form for $t>t'_{12}$,
we may write:
\ba
 \psi^{tr}_{\Delta k_x}(x,t) & = & \frac{1}{\sqrt{2\pi}} \ \!
  e^{- i\Delta k_x^2 \hbar t/2m} \nonumber \\
  & & \times \int_{-\infty}^\infty dk_x c(k_x - \Delta k_x)
   e^{-i k_x^2 \hbar t/2m} e^{ik_x (x + \Delta x_0)} .
 \label{eq17}
\ea
By changing now the integration variable $k'_x = k_x - \Delta k_x$ and
using the relation $\hbar t/m = y/k$, (\ref{eq17}) transforms into
\ba
 \psi^{tr}_{\Delta k_x}(x,y) & = & \frac{1}{\sqrt{2\pi}}\ \!
  e^{i\Delta k_x (x + \Delta x_0) - i\Delta k_x^2 y/k} \nonumber \\
 & & \times \int_{-\infty}^\infty dk'_x c(k'_x)
   e^{-i {k'}^2_x y/2k} e^{ik'_x (x + \Delta x_0 - y\Delta k_x/k)} .
 \label{eq18}
\ea
Then, after multiplying (\ref{eq18}) by $e^{iky}$, we obtain the
space-dependent wave function which is the continuation of (\ref{eq3})
for $y > y'_{12}$, i.e.,
\be
 \psi_{\Delta k_x} (x,y) = e^{iky} \psi^{tr}_{\Delta k_x} (x,y) .
 \label{eq19}
\ee
In analogy to the approximation (\ref{eq6}) for (\ref{eq3}) and
(\ref{eq5}), the wave function (\ref{eq19}) can also be approximated
in the far field by the simpler form,
\ba
 \psi_{\Delta k_x} (x,y) & = & \sqrt{\frac{k}{y}} \ \! e^{iky}
  e^{-i\pi/4} e^{-i\Delta k_x^2 y/2k} \nonumber \\
 & & \times
  e^{ik(x + \Delta x_0)^2/2y} c[(k(x+\Delta x_0)/y - \Delta k_x) .
 \label{eq20}
\ea
Assuming that the beam incident to the first grating is a plane wave
that illuminates $n$ slits, from (\ref{eq4}) we find:
\be
 c(k_x) = \frac{\sqrt{2}}{\sqrt{\pi n}\delta}
  \frac{\sin(k_x \delta/2)}{k_x}
  \frac{\sin(k_x dn/2)}{\sin(k_x d/2)} ,
 \label{eq21}
\ee
where $d$ is the grating period and $\delta$ is the slit width.

\begin{figure}
 \begin{center}
 \epsfxsize=13cm {\epsfbox{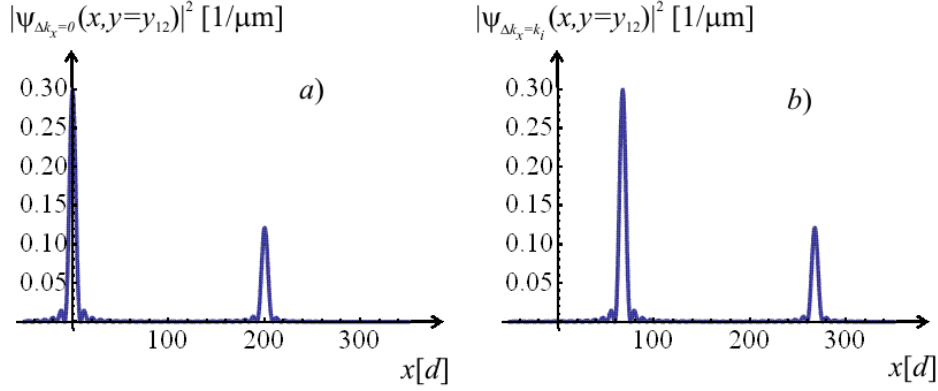}}
 \caption{\label{fig2}
  The function $|\psi_{\Delta k_x}(x,y=y_{12})|^2$ when the laser is
  off (a), with $\Delta k_x = 0$, and when the laser is on (b), with
  $y'_{12} = 5kd/8k_i$ and $\Delta k_x = k_i$.
  The parameters considered are: $v = 1400$~m/s,
  $k = m_{\rm Na} v/\hbar = 5.09067 \cdot 10^{11}$~m$^{-1}$,
  $k_i = 2\pi/(589~{\rm nm}) = 1.06675 \cdot 10^7$~m$^{-1}$,
  $y_{12} = y_{23} = 0.65$~m, $d = 2 \cdot 10^{-7}$~m,
  $\delta = 1 \cdot 10^{-7}$~m and $n = 24$.}
 \end{center}
\end{figure}

The wave function $\psi_{\Delta k_x}(x,y=y_{12})$ that reaches the
second grating has two narrow maxima, each one covering several slits.
The square modulus of this function is shown in figure~\ref{fig2}a for
the laser off and in figure~\ref{fig2}b for the laser on.

\section{The wave function behind the second grating}
 \label{sec4}

In order to determine the wave function behind the second grating, it
is convenient to apply the form (\ref{eq5}) of the atomic wave function.
Thus, we have
\be
 \psi(x,y) = \sqrt{\frac{k}{2\pi y}} \ \! e^{-i\pi/4} e^{iky}
  \int_{-\infty}^\infty dx' \psi(x',y_{12}^{+0})
  e^{ik(x-x')^2/2y} , \qquad y > y_{12} ,
 \label{eq22}
\ee
where $\psi(x',y_{12}^{+0})$ is the wave function just after the second
grating.

\begin{figure}
 \begin{center}
 \epsfxsize=13cm {\epsfbox{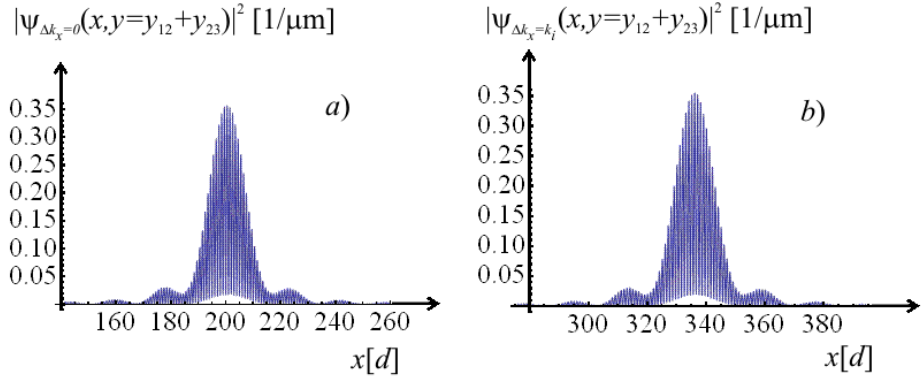}}
 \caption{\label{fig3}
  The function $|\psi_{\Delta k_x}(x,y=y_{12} + y_{23})|^2$ when the
  laser is off (a), with $\Delta k_x = 0$, and when the laser is on
  (b), with $y'_{12} = 5kd/8k_i$ and $\Delta k_x = k_i$.
  The parameters considered are the same as in figure \ref{fig2}.
  The period of the fast oscillations observed is the same as the
  grating period.}
 \end{center}
\end{figure}

If the laser is off ($\Delta k_x = 0$), the wave function does not
depend on $y'_{12}$.
We then find that the square modulus of the wave function incident
to the third grating has the form shown in figure~\ref{fig3}a: it
oscillates with period $d$.
If the laser is turned on, the function has again the same form, but it
undergoes a shift along the $x$-axis (see figure~\ref{fig3}b) for an
amount that depends on $\Delta k_x$.

\section{Transmission through the third grating}
 \label{sec5}

In the experiment of Chapman \etal~\cite{chapman1} the
corresponding patterns were obtained by counting the number of atoms
transmitted through the third grating.
So, in order to compare the above analytical results with experimental
data, it is necessary to evaluate the number of transmitted atoms
through the third grating for various values of its lateral shift
$\Delta x_3$.
The transmission is evaluated by integrating first the intensity in
the region of the first maximum (i.e., in the range of $x$ shown in
figure~\ref{eq4}) for fixed values of the lateral shift and transferred
impulses $\Delta k_x$ to the atom during the photon scattering, i.e.,
\be
 T(y'_{12}, \Delta k_x, \Delta x_3) = \int_{\rm slits}
  \left\arrowvert \psi_{\Delta k_x} (x, y = y_{12} + y_{23})
  \right\arrowvert^2 dx .
 \label{eq23}
\ee
Numerical results which we have obtained for different
values of $y'_{12}$ and $\Delta k_x$ show that the function
$T(y'_{12}, \Delta k_x, \Delta x_3)$ has the following simple
periodic form:
\be
 T(y'_{12}, \Delta k_x, \Delta x_3) = a
  + b \cos (2\pi \Delta x_3 /d + d_p \Delta k_x) ,
 \label{eq24}
\ee
where $a$ and $b$ are constants which do not depend on $y'_{12}$ and
$\Delta k_x$, and the quantity
\be
 d_p = (2\pi/kd) y'_{12}
 \label{eq25}
\ee
is the distance between the paths (the lines of maxima of the atomic
wave function) at the place of scattering with a photon.

\begin{figure}
 \begin{center}
 \epsfxsize=8cm {\epsfbox{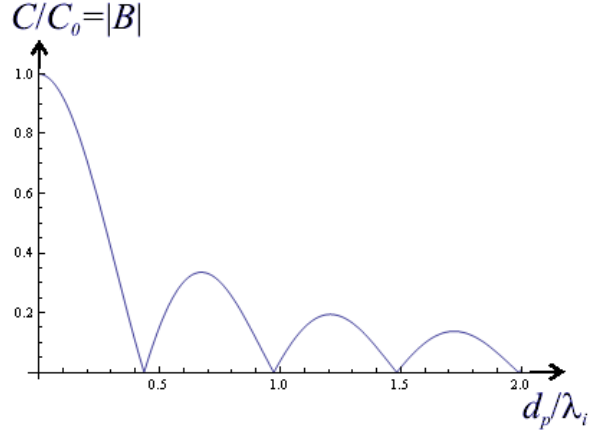}}
 \caption{\label{fig4}
  Relative contrast as a function of $d_p/\lambda_i$.}
 \end{center}
\end{figure}

Next, we have to integrate over all possible values of the transferred
momentum taking into account the probability distribution of the
transferred momentum, $P_1(\Delta k_x)$.
As shown by Mandel and Wolf \cite{wolf}, this distribution is given by
\be
 P_1 (\Delta k_x) = \frac{3}{8k_i}
  \left[ 1 + \left(1 - \frac{\Delta k_x}{k_i} \right)^2 \right] .
 \label{eq26}
\ee
Consequently,
\ba
 T(y'_{12}, \Delta x_3) & = &
  \int_0^{2k_i} d(\Delta k_x) P_1 (\Delta k_x)
  T(y'_{12}, \Delta k_x, \Delta x_3) \nonumber \\
 & = &
  \int_0^{2k_i} d(\Delta k_x) \ \! \frac{3}{8k_i}
  \left[ 1 + \left( 1 - \frac{\Delta k_x}{k_i} \right)^2 \right]
  \nonumber \\
 & & \times (a + b \cos (2\pi \Delta x_3 /d + d_p \Delta k_x)) ,
 \label{eq27}
\ea
After analytical integration of (\ref{eq27}), we obtain
\be
 T(y'_{12}, \Delta x_3) = a
  + b B \cos (2\pi \Delta x_3/d + d_p k_i ) ,
 \label{eq28}
\ee
where
\be
 B = \frac{3}{4\pi} \frac{\lambda_i}{d_p} \left[
  \left( 1 - \frac{1}{(2\pi)^2} \frac{\lambda_i^2}{d_p^2} \right)
    \sin (2\pi d_p/\lambda_i ) + \frac{1}{2\pi} \frac{\lambda_i}{d_p}
    \cos (2\pi d_p/\lambda_i ) \right] .
 \label{eq29}
\ee
As is apparent from (\ref{eq28}), the contrast when the laser is off
and on is determined by the quantities $a$, $b$ and $B$, as
\be
 C_0 = \left\arrowvert \frac{b}{c} \right\arrowvert , \qquad
  C = \frac{T_{\rm max} - T_{\rm min}}{T_{\rm max} + T_{\rm min}}
    = \left\arrowvert \frac{b}{a} \ \! B \right\arrowvert ,
 \label{eq30}
\ee
with the relative contrast being
\be
 \frac{C}{C_0} = |B| .
 \label{eq31}
\ee
The relative contrast displayed in figure~\ref{fig4} is an analytic
function of the ratio $d_p/\lambda_i$.

\section{Conclusions}
 \label{sec6}

Our description and explanation of the experiment by Chapman \etal
\cite{chapman1,chapman2} is based on the assumption that there
is a wave associated with an atom.
The evolution of the wave is determined by the Schr\"odinger equation,
the boundary conditions imposed by the gratings and the interaction
between the atom and a photon.
As shown here, an initial harmonic atomic wave is transformed by the
first grating into a wave with narrow maxima at the points along and
in close vicinity of three particular paths (though only two of them
are of relevance in this experiment) and negligible values at any other
point.
The two maxima move together; in other words, the wave is coherent.
At the grating, the particle associated with the wave acquires randomly
a new value for its momentum which directs the particle towards one of
the paths along which it moves following the time evolution of a wave
field.
The photon scattering that takes place between the first and second
gratings causes the change of the atomic transverse momentum.
Consequently, the atomic wave function is shifted along the $x$-axis,
but without destroying the coherence, and the contrast of the
transmission function will not depend neither on the point of
scattering nor on the photon wavelength.

The dependence of the transmission on the ratio $d_p/\lambda_i$ is
obtained after integrating over all possible values of transferred
momenta.
In this explanation, wave and particle properties are compatible since
both are present and play a role. Within the model presented here, the
behavior of contrast can be explained for all values of
$d_p/\lambda_i$.
Moreover, the problem of explaining the so-called revivals of the
coherence after it was ``lost'' at $d_p/\lambda_i \approx 0.5$ does
not appear, as required by complementarity.


\ack

MB is grateful to D Pritchard for suggesting the study of this problem
and M Benedict for fruitful discussions in Szeged.
MD, DA and MB acknowledge support from the Ministry of Science of
Serbia under Project `Quantum and Optical Interferometry', number
141003.
ASS acknowledges support from the Ministerio de Ciencia e Innovaci\'on
(Spain) under project number FIS2007-62006 and the Consejo Superior de
Investigaciones Cient\'{\i}ficas through a JAE-Doc Contract.


\end{document}